\documentclass[preprint2]{aastex631}
\usepackage{amsmath}
\usepackage{ulem}
\usepackage{hyperref}
\setcitestyle{notesep={ }}
\usepackage{bm}

\graphicspath{{./}{figures/}}

\definecolor{ao(english)}{rgb}{0.0, 0.5, 0.0}

\begin{document}

\title{Alfv\'en Wave Mode Conversion in Neutron Star Magnetospheres: A Semi-analytic Approach}

\correspondingauthor{Alexander Y. Chen}
\email{cyuran@wustl.edu}

\author[0000-0002-4738-1168]{Alexander Y. Chen}
\affil{Physics Department and McDonnell Center for the Space Sciences, Washington University in St. Louis; MO, 63130, USA}

\author[0000-0002-0108-4774]{Yajie Yuan}
\affil{Physics Department and McDonnell Center for the Space Sciences, Washington University in St. Louis; MO, 63130, USA}

\author[0009-0008-1825-6043]{Dominic Bernardi}
\affil{Physics Department and McDonnell Center for the Space Sciences, Washington University in St. Louis; MO, 63130, USA}

\begin{abstract}
    We write down the force-free electrodynamics (FFE) equations in dipole coordinates, and solve for normal modes corresponding to Alfv\'enic perturbations in the magnetosphere of a neutron star. We show that a single Alfv\'en wave propagating on dipole field lines spontaneously sources a fast magnetosonic (fms) wave at the next order in the perturbation expansion, without needing 3-wave interaction. The frequency of the sourced fms wave is twice the original Alfv\'en wave frequency, and the wave propagates spherically outwards. The properties of the outgoing fms wave can be computed exactly using the usual devices of classical electrodynamics. We extend the calculation to the closed zone of a rotating neutron star magnetosphere, and show that the Alfv\'en wave also sources a spherical fms wave but at the same frequency as the primary Alfv\'en wave.
\end{abstract}


\section{Introduction} \label{sec:intro}

Nonlinear wave phenomena in the magnetospheres of neutron stars have recently
garnered significant attention in the community. It was demonstrated
through first-principles force-free simulations that small-amplitude Alfv\'en
waves can spontaneously convert to fast magnetosonic (fms)
waves~\citep{2021ApJ...908..176Y}, and large-amplitude Alfv\'en waves can become
nonlinear and break the background magnetic field lines, launching a
relativistic ejecta~\citep{2020ApJ...900L..21Y,2022ApJ...933..174Y}. This
relativistic ejecta can potentially launch shocks at large radii and power Fast
Radio Bursts
(FRBs)~\citep[e.g.][]{2019MNRAS.485.4091M,2019MNRAS.485.3816P,2021PhRvL.127c5101S}.
Alfv\'en waves propagating on curved magnetic field lines also suffer from
transverse steepening (also called dephasing by \citet{2020ApJ...897..173B}), which leads to
spontaneous increase in $k_{\perp}$ as the wave propagates and may
lead to charge separation~\citep{2022ApJ...929...31C}.

In addition, recent simulations also studied nonlinear interactions between
colliding Alfv\'en waves. For oppositely polarized Alfv\'en waves, their
magnetic field components may cancel, leaving a region with
$E > B$~\citep{2019ApJ...881...13L,2021ApJ...915..101L}. This can lead to fast
dissipation of the colliding Alfv\'en wave, creating an evanescent current sheet
that can give rise to nonthermal particle acceleration.
\citet{2021JPlPh..87f9014T} and \citet{2022PhRvL.128g5101N} studied the collision between two oblique relativistic
Alfv\'en waves, and found that this process generally leads to a turbulent cascade. In particular, \citet{2022PhRvL.128g5101N} showed that charge-starvation may
occur during such a cascade which leads to a distinct dissipation mechanism. 

Fast magnetosonic waves, on the other hand, can steepen into very strong shocks
when they become nonlinear~\citep{2022arXiv221013506C,2023ApJ...959...34B}. This
mechanism can in principle provide strong constraints on the location where FRBs
are produced, preventing FRBs emitted near the star to escape from the closed
zone of the magnetosphere~\citep{2023arXiv230712182B}. Even in regimes where the
fms wave does not steepen, it may still suffer from significantly enhanced
scattering cross section with charge particles, leading to strong energy loss
and limiting the optically thin region to a small cone near the magnetic
axis~\citep{2022PhRvL.128y5003B,2022MNRAS.515.2020Q}. Pulses of fms waves can
interact with the equatorial current sheet near the light cylinder, leading to
fast reconnection that can also potentially produce FRB
signals~\citep{2020ApJ...897....1L,2022ApJ...932L..20M,2023RAA....23c5010W}. The
fms waves may spontaneously decay into Alfv\'en waves through 3-wave interaction
and their energy transferred to small scales
directly~\citep{2023ApJ...957..102G}. It was also recently proposed that fms
waves may also scatter off magnetospheric plasma via inverse Compton scattering,
which can be another potential channel for producing FRB signals from within the
magnetosphere~(Qu et al. in prep.).

The nonlinear phenomena associated with Alfv\'en waves and fms waves are often
analyzed in the WKB framework, where the wavelength is taken to be much smaller
than the global length scale $L$ (e.g.\ the length of the magnetic field line on
which they
propagate)~\citep{2020ApJ...897..173B,2021ApJ...908..176Y,2023ApJ...957..102G}.
Similarly, in many of the simulation works on wave interactions cited above, the
curvature of the background magnetic field is often ignored. However, for
Alfv\'en waves launched by star quakes, their characteristic angular frequency
can be ${\sim}10\,\mathrm{kHz}$~\citep[e.g.][]{2020ApJ...897..173B}, leading to
a macroscopic wavelength $\lambda_{A} \sim 3\times 10^{6}\,\mathrm{cm}$. As a
result, the WKB assumption of $kL \gg 1$ may not always be reasonable especially
for field lines close to the star.

We believe a solution of the Alfv\'en wave in a dipole background magnetic field
can help facilitate the study of various nonlinear phenomena in the neutron star
magnetosphere. To our knowledge, such a solution does not explicitly exist in
the literature, and it can potentially be used to elucidate the nature of
nonlinear wave mode conversion in the magnetosphere. In this work, we attempt to
construct such a solution under an asymptotic expansion in wave amplitude,
without resorting to a WKB approximation. Then, we study the higher order
effects caused by the propagation of this Alfv\'en wave. In particular, we focus
on its conversion into fms waves within the neutron star magnetosphere.

This paper is organized into three main parts. In
Section~\ref{sec:non-rotating}, we derive the equation governing Alfv\'enic
normal modes in dipole geometry. We solve the equation numerically and compare
it with previously suggested approximate results. In
Section~\ref{sec:second-order}, we expand the field quantities to second order
and show that the primary Alfv\'en wave spontaneously sources a second-order fms
wave due to the geometry of dipole field lines. We use the numerical solution
from Section~\ref{sec:non-rotating} to compute the conversion rate to fms waves
and discuss parameter scaling. In Section~\ref{sec:rotating}, we carry out a
two-scale asymptotic expansion in both electromagnetic perturbation amplitude
and rotation velocity. We show that a similar conversion to fms waves now
happens at first order in both perturbation amplitude and rotation. Finally in
Section~\ref{sec:discussion} we discuss potential implications of this work and future directions.

\section{Alfv\'en Waves in a Non-rotating Dipole Magnetosphere} \label{sec:non-rotating}

In a dipole background magnetic field, the magnetic flux coordinates align with the
dipole field lines. There are multiple ways to construct an orthogonal coordinate system
out of this geometry~\citep[e.g.][]{2006CG.....32..265K}, but we will adopt the standard dipole coordinates for this paper:
\begin{equation}
    \label{eq:dipole-coord}
    q = \frac{\cos\theta}{r^{2}},\quad p = \frac{r}{\sin^{2}\theta},\quad \phi = \phi,
\end{equation}
where $q$ varies along the dipole field line, and $p$ varies perpendicular to
the field line. Note that $q$ increases from the southern hemisphere to the
northern hemisphere due to the factor of $\cos\theta$. The variable $p$ labels
individual field lines and marks their maximum extent,
$p = r_\mathrm{max}$. We also define an auxiliary quantity
$\delta = \sqrt{1 + 3\cos^{2}\theta}$. A number of useful expressions including
vector calculus identities are listed in Appendix~\ref{sec:app-dipole-coord}.

The force-free equations read~\citep[see e.g.][]{1999astro.ph..2288G}:
\begin{equation}
  \label{eq:ffe}
  \begin{split}
  \partial_{t} \bm{E} &= c\nabla\times \bm{B} - 4\pi\bm{j} \\
  \partial_{t} \bm{B} &= -c\nabla\times \bm{E} \\
  \bm{j} &= \frac{c}{4\pi}\frac{\bm{B}\cdot \nabla\times \bm{B} - \bm{E}\cdot \nabla\times \bm{E}}{B^{2}}\bm{B} \\
    &\quad + \frac{c}{4\pi}\nabla\cdot \bm{E}\frac{\bm{E}\times \bm{B}}{B^{2}}.
  \end{split}
\end{equation}
We decompose the electromagnetic fields in an asymptotic expansion in
$\delta B/B_{0}$ where $B_{0}$ is the strength of the background magnetic field:
\begin{equation}
    \label{eq:asymp-expansion-1}
    \begin{split}
    \bm{B} &= \bm{B}^{(0)} + \bm{B}^{(1)} + \bm{B}^{(2)} + \dots \\
    \bm{E} &= \bm{E}^{(0)} + \bm{E}^{(1)} + \bm{E}^{(2)} + \dots \\
    \end{split}
\end{equation}
where $\bm{B}^{(0)} = \bm{B}_{0} \propto r^{-3}\hat{\bm{q}}$ is the dipole
background, and $\bm{E}^{(0)} = 0$. We seek wave solutions for the first
order fields $\bm{E}^{(1)}$ and $\bm{B}^{(1)}$.

In axisymmetry, Alfv\'en waves have magnetic perturbation in the
$\hat{\boldsymbol{\phi}}$ direction. They are ducted along the field
line, and their amplitude scales as approximately $r^{-3/2}$. Therefore,
we look for 2D solutions of the following general form:
\begin{equation}
    \label{eq:B-ansatz-f}
    \bm{B}^{(1)} = r^{-3/2}e^{-i\omega t}f(p, q)\hat{\boldsymbol{\phi}}.
\end{equation}
and our goal is to determine the unknown function $f(p, q)$. This is equivalent
to taking a Fourier transform in time on the linearized equations, and we are
seeking the normal modes of electromagnetic oscillations in the dipole geometry.
\citet{2022ApJ...929...31C} wrote down an ansatz for the Alfv\'en wave solution
that is equivalent to $f(p, q) = \exp(\Phi_{0}(p) + ik_{\parallel}s)$ where $s$
is the field line length measured from the launching footpoint, and $\Phi_{0}$
is the initial wave phase that may be different on different field lines. In the
following, we will derive the equation
governing $f(p, q)$ and check whether this ansatz is correct.

The first order force-free current density is given by only a single term
since the background $\bm{E}^{(0)}$ vanishes and $\nabla\times \bm{B}^{(0)} = 0$:
\begin{equation}
    \label{eq:j1}
    \bm{j}^{(1)} = \frac{c}{4\pi}\frac{\bm{B}^{(0)}\cdot \nabla\times \bm{B}^{(1)}}{B_{0}^{2}}\bm{B}^{(0)}.
\end{equation}
Plugging the Alfv\'en wave ansatz and this first order current into the Maxwell
equations, it can be seen that $\bm{j}^{(1)}$ cancels the $\hat{\bm{q}}$
component of $\nabla\times \bm{B}^{(1)}$, leaving only $\hat{\bm{p}}$
component for $\partial \bm{E}^{(1)}/\partial t$:
\begin{equation}
    \label{eq:E1}
    \begin{split}
    \partial_{t} \bm{E}^{(1)} &= c\nabla\times \bm{B}^{(1)} - 4\pi\bm{j}^{(1)} \\
      &= c\hat{\bm{p}}\left[-\frac{\delta}{r^{3}}\frac{\partial B_{\phi}^{(1)}}{\partial q} + \frac{3\cos\theta}{r\delta}B_{\phi}^{(1)}\right] \\
      &= c\hat{\bm{p}}\left(-\frac{\delta}{r^{3}}\frac{\partial_{q}f}{f}B_{\phi}^{(1)}\right).
    \end{split}
\end{equation}
The factor of $r^{-3/2}$ in Equation~\eqref{eq:B-ansatz-f} leads to a
cancellation of the second term on the second line, resulting in a relatively
simple expression. If we assume $e^{-i\omega t}$ time dependence for
$\bm{E}^{(1)}$ as well, then:
\begin{equation}
    \label{eq:Ep}
    E_{p}^{(1)} = -\frac{ic}{\omega}\frac{\delta}{r^{3}}\frac{\partial_{q}f}{f}B_{\phi}^{(1)}.
\end{equation}

Given that $\bm{E}$ is only nonzero in the $\hat{\bm{p}}$ direction, the other Maxwell equation can be simplified too:
\begin{equation}
    \label{eq:B1}
    \begin{split}
        \partial_{t} \bm{B}^{(1)} &= -c\nabla \times \bm{E}^{(1)}\\
        &= -c\hat{\boldsymbol{\phi}}\left[\frac{\delta}{r^{3}}\frac{\partial E_{p}^{(1)}}{\partial q} - \frac{6\cos\theta}{r\delta^{3}}(1 + \cos^{2}\theta)E_{p}^{(1)}\right]
    \end{split}
\end{equation}

Combining Equations~\eqref{eq:E1} and \eqref{eq:B1}, we can write down a second-order differential equation for the unknown function $f(p, q)$:
\begin{equation}
    \label{eq:f-equation}
    \partial_{q}(r^{-3}\partial_{q} f) = -\left(\frac{\omega}{c}\right)^{2}\frac{r^{6}}{\delta^{2}}f.
\end{equation}
In particular, $f \propto e^{ik_{\parallel}s}$ turns out not to be a solution to
Equation~\eqref{eq:f-equation}. Plugging it in and using
$\partial s/\partial q = r^{3}/\delta$, we are left with
$ik_{\parallel}f\partial_{q}(1/\delta) = 0$, which does not hold for general
$k_{\parallel}$.

Note that Equation~\eqref{eq:f-equation} has no $p$ derivatives, which means
that the wave profile is essentially decoupled between field lines. The wave on
each field line propagates independently. \citet{2020ApJ...897..173B} also
explicitly used this property when evolving the Alfv\'en waves in the magnetosphere. We can therefore specialize to a single magnetic field line with
constant $p$, i.e.\ $r = p \sin^{2}\theta$, and interpret the equation as an ODE.
It is also possible to change the variable from $q$ to $\mu = \cos\theta$ to
further simplify the equation:
\begin{equation}
    \label{eq:f-equation-mu}
    \frac{d}{d\mu}\left(\frac{1}{1 + 3\mu^{2}}\frac{df}{d\mu}\right) = -\kappa^{2}f,
\end{equation}
where we have defined a dimensionless wave number $\kappa = \omega p/c$ that can
be understood as the ratio between $p$ and the Alfv\'en wavelength
$\lambda$. This is the differential equation governing the spatial part of
Alfv\'en waves in dipole geometry.

\begin{figure}[h]
    \centering
    \includegraphics[width=0.47\textwidth]{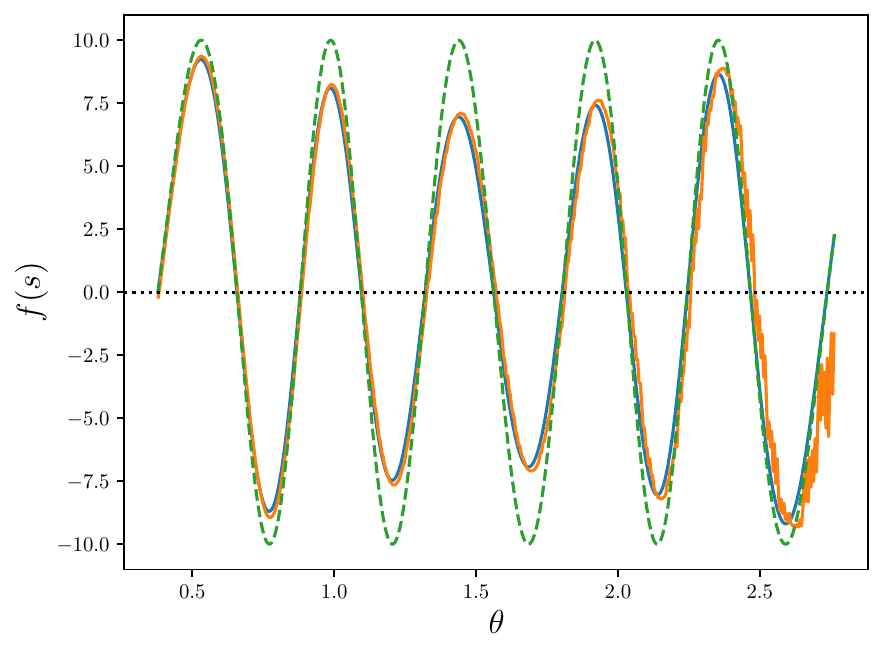}
    \caption{Blue solid curve shows a numerical solution of
      Equation~\eqref{eq:f-equation-mu} on a specific field line
      $p \approx 7.3R_{*}$ and $\kappa = \omega p/c \approx 12.74$. The
      numerical solution is obtained using an adaptive Dormand-Prince RK45
      algorithm. The orange curve shows an interpolated result from a 2D axisymmetric
      force-free simulation (Bernardi et al. in prep.), and the dashed green
      curve shows the reference function $\sin(k_{\parallel}s)$, where
      $k_{\parallel} = \omega/c$. The interpolated FFE simulation results
      deteriorates at large $\theta$ due to significant dephasing when the wave
      reaches the opposite footpoint.}
    \label{fig:f-solution}
\end{figure}

Figure~\ref{fig:f-solution} shows a numerical solution for
Equation~\eqref{eq:f-equation-mu} on a specific field line, in comparison with
the result from a 2D axisymmetric force-free simulation and the reference
solution $\sin(k_{\parallel}s)$. One can see that the solution exactly replicate
the results from our 2D force-free simulation. The solution is wave-like with
the same wavelength $\lambda = 2\pi/k_{\parallel}$, but there is an amplitude
modulation that decreases near the equatorial plane compared to
$\sin(k_{\parallel}s)$. This amplitude modulation can be interpreted as a correction on the $r^{-3/2}$ amplitude dependence, and is effectively equivalent to the evolution of a WKB mode as $k_{\parallel}\to\infty$.

Since Equation~\eqref{eq:f-equation-mu} is a second order equation, it admits
two linearly independent solutions. In general, we write these two solutions
as $f_{1}$ and $f_{2}$, which can be chosen to be the analog of $\cos kx$ and
$\sin kx$. The full time-dependent solution can therefore be constructed as:
\begin{equation}
    \label{eq:Bphi-time-dependent}
    B_{\phi}^{(1)}(t) = r^{-3/2}(f_{1} + if_{2})e^{-i\omega t}.
\end{equation}
It is understood that the Alfv\'en wave is the real part of this solution, but
the imaginary part is necessary when using Equation~\eqref{eq:Ep} to compute the
wave electric field. Since $f_{1}$ and $f_{2}$ are completely determined by
solving a relatively simple ODE, this way of constructing the time-dependent
solution is computationally much more efficient than evolving a time-dependent
PDE, such as done by~\citet{2020ApJ...897..173B}. A more detailed description of
how $f_{1}$ and $f_{2}$ are obtained is deferred to
Appendix~\ref{sec:app-alfven-solution}.

\section{Second Order: Conversion to FMS Waves} \label{sec:second-order}

Now that we have a solution of the first order normal mode corresponding to
general small-amplitude Alfv\'en waves in dipole geometry, we can evaluate the
second order field quantities $\bm{B}^{(2)}$, $\bm{E}^{(2)}$, and $\bm{j}^{(2)}$. This
will allow us to study nonlinear wave conversion phenomena directly. In the rest
of this paper, we will focus on the spontaneous evolution of a single
small-amplitude Alfv\'en wave, and defer the more complicated case of
interaction between different waves to a future study.

\subsection{Formalism}
\label{sec:formalism}

Assuming that the solution at first order only contains a single Alfv\'en
wave normal mode described by Equation~\eqref{eq:B-ansatz-f}, the second order
current has only three terms:
\begin{equation}
    \label{eq:j2}
    \begin{split}
    \bm{j}^{(2)} &= \frac{c}{4\pi}\left[\frac{\bm{B}^{(0)}\cdot\nabla\times \bm{B}^{(2)}}{B_{0}^{2}}\bm{B}^{(0)}\right. \\
                       &\quad + \frac{\bm{B}^{(0)}\cdot\nabla\times \bm{B}^{(1)}}{B_{0}^{2}}\bm{B}^{(1)} \\
                       &\quad + \left.\nabla\cdot \bm{E}^{(1)}\frac{\bm{E}^{(1)}\times \bm{B}^{(0)}}{B_{0}^{2}}\right].
    \end{split}
\end{equation}
The term $\bm{B}^{(1)}\cdot \nabla\times \bm{B}^{(1)}$ vanishes
due to $\nabla\times \bm{B}^{(1)}$ being in the $\hat{\bm{p}}$
direction. Similarly $\bm{E}^{(1)}\cdot \nabla\times \bm{E}^{(1)}$
vanishes due to $\nabla\times \bm{E}^{(1)}$ only present in the
$\hat{\boldsymbol{\phi}}$ direction. The $B^{2}$ term in the denominators
expands into
$B^{2} = B_{0}^{2} + 2\bm{B}^{(1)}\cdot \bm{B}^{(0)} + 2\bm{B}^{(2)}\cdot \bm{B}^{(0)} + \left(\bm{B}^{(1)}\right)^{2} + \dots$,
but since $\bm{B}^{(1)}\cdot \bm{B}^{(0)} = 0$, the denominator has no
first order corrections, therefore the extra terms only come in at $\bm{j}^{(3)}$
or above.

The three terms in Equation~\eqref{eq:j2} have well-defined directions. The
first term is an Alfv\'en wave current along the background field, with exactly
the same structure as $\bm{j}^{(1)}$. It leads to a correction to our first order
Alfv\'en wave solution without changing its polarization, therefore the solution is still fully described by what was discussed in Section~\ref{sec:non-rotating}. The amplitude and frequency of this second order Alfv\'en wave is apparently unrelated to the first order wave, and need to be determined by additional constraints such as energy conservation.

The second and third terms in Equation~\eqref{eq:j2} are both along the
$\hat{\boldsymbol{\phi}}$ direction, so we can group them and denote them as
$\bm{j}_{\phi}^{(2)}$. Note that $\bm{j}_{\phi}^{(2)}$ only depends on the first order Alfv\'en
wave solution, therefore it only acts as a source term for the second order
wave. The equations for second-order perturbations sourced by this current are:
\begin{equation}
    \label{eq:2nd-order-fms}
    \begin{split}
    \partial_{t} \bm{E}^{(2)} &= c\nabla\times \bm{B}^{(2)} - 4\pi \bm{j}_{\phi}^{(2)}, \\
    \partial_{t} \bm{B}^{(2)} &= -c\nabla \times \bm{E}^{(2)}.
    \end{split}
\end{equation}
Equations~\eqref{eq:2nd-order-fms} are identical to the vacuum linear Maxwell
equations with an external current. The direction of $\bm{j}$ suggests that
$\bm{E}^{(2)}$ is only along the $\hat{\boldsymbol{\phi}}$ direction, and the
structure of the equations suggests wave solutions that are degenerate with
vacuum electromagnetic waves. Both of these properties point to a fms wave.
Therefore, we can interpret the first order Alfv\'en wave acting as an antenna,
producing time-varying $\bm{j}_{\phi}^{(2)}$ that sources a second order fms
wave. Since both terms in $\bm{j}_{\phi}^{(2)}$ are second order in
$\bm{B}^{(1)}$ or $\bm{E}^{(1)}$, the oscillation frequency $\omega_{2}$ of
$\bm{j}_{\phi}^{(2)}$ is twice the original Alfv\'en wave frequency,
$\omega_{2} = 2\omega$. The source fms wave is expected to have the same
frequency as $\bm{j}_{\phi}^{(2)}$. More quantitative discussion of the
frequency structure will be presented in Section~\ref{sec:quant}.

In Cartesian geometry, the second and third terms in Equation~\eqref{eq:j2}
exactly cancel each other, therefore there is no spontaneous production of
second order fms waves. Similarly, under the usual WKB assumption where
$kL\gg 1$, the background magnetic field is locally constant and the first order solution behaves like a plane wave, therefore
$j_{\phi}$ also vanishes. However, in general curvilinear geometry when the
curvature radius is not too much larger than the wavelength, there is no
guarantee that these two terms cancel. As a result, we argue that when a first
order Alfv\'en wave propagates on curved field lines, it should spontaneously
source a second order fms wave by acting as an antenna. The resulting fms wave
has double the original Alfv\'en wave frequency. This is a geometrical nonlinear
effect that is not captured in the WKB approximation. In the limit of $kL\gg 1$ however,
we do expect the spontaneous conversion to fms waves to become negligible, as the
wave solution approaches the usual plane wave solution in a uniform magnetic field.

The analysis in this section differs from existing literature based on 3-wave
interaction~\citep[see e.g.][]{2019MNRAS.483.1731L,2021ApJ...908..176Y} in two
ways. Firstly, if one attributes the generation of fms waves from the primary
Alfv\'en wave to some form of 3-wave interaction, then the natural question is
what is the Alfv\'en wave interacting with. \citet{2021ApJ...908..176Y}
hypothesized that it is the interaction between the primary Alfv\'en wave and a
back-propagating Alfv\'en wave that produces the outgoing fms wave. However,
this picture does not make a definitive prediction about the frequency of the
outgoing wave, since there is not much information to constrain the
back-propagating Alfv\'en wave.
Secondly, 3-wave interaction typically requires the kinematic constraint of
$\bm{k}' = \bm{k}_{1} + \bm{k}_{2}$. However, a quasi-isotropic
outgoing spherical fms wave would need to involve $\bm{k}'$ of all directions,
necessitating a wide range of combinations of interacting waves $\bm{k}_{1}$
and $\bm{k}_{2}$ that are difficult to extract from the propagating, spatially extended primary Alfv\'en wave.

The analysis outlined above shows that a primary Alfv\'en wave spontaneously
converting to a second order fms wave is the result of nonlinearity on curved
magnetic field lines. This analysis makes a definitive prediction that the
frequency of the outgoing fms wave is twice the Alfv\'en wave frequency, which
agrees with the simulations performed by \citet{2021ApJ...908..176Y}. This
picture also allows us to quantitatively compute the properties of the resulting
fms wave using classical electrodynamics. We present such a calculation in Section~\ref{sec:quant}.

\subsection{Quantitative Calculations}
\label{sec:quant}

The key to computing the spontaneous conversion of a single Alfv\'en wave to fms
waves is the second order current density $j_{\phi}^{(2)}$ in
Equation~\eqref{eq:j2}. Once we find the Fourier components of $j_{\phi}^{(2)}$,
we can use it to compute the vector potential $A_{\phi}^{(2)}$, which can be
used to compute the electric and magnetic fields of the fms wave. The Fourier
components of the vector potential $A_{\phi}^{(2)}$ can also be used to directly
compute the power and the angular dependence of the outgoing wave.

In component form, the equation for $j_{\phi}^{(2)}$ can be written as:
\begin{equation}
    \label{eq:jphi}
    \begin{split}
      j_{\phi}^{(2)} &= \frac{c}{4\pi}\left[ \frac{B_{\phi}^{(1)}}{B_{0}}\left(\frac{\delta}{\sin^{3}\theta}\frac{\partial B_{\phi}^{(1)}}{\partial p} + \frac{1 - 3\cos^{2}\theta}{r\delta \sin\theta}B_{\phi}^{(1)}\right)\right. \\
      &- \left.\frac{E_{p}^{(1)}}{B_{0}}\left(\frac{\delta}{\sin^{3}\theta}\frac{\partial E^{(1)}_{p}}{\partial p} + \frac{4(1 - 3\cos^{4}\theta)}{r\delta^{3}\sin\theta}E^{(1)}_{p}\right)\right] \\
      &= \frac{c}{4\pi}\left[\frac{B_{\phi}^{(1)}}{B_{0}}\hat{D}_{B}B_{\phi}^{(1)} - \frac{E_{p}^{(1)}}{B_{0}}\hat{D}_{E}E_{p}^{(1)}\right],
    \end{split}
\end{equation}
where we have introduced two operators $\hat{D}_{B}$ and $\hat{D}_{E}$ to
facilitate subsequent calculations. As evident from Equation~\eqref{eq:jphi},
the second order current depends on the partial derivative of the wave fields
across field lines, $\partial_{p}$. Since the waves on each field line propagate
independently, this derivative is determined by the initial phase gradient
of the wave perpendicular to the field lines, as well as the phase difference
built up due to different field line lengths~\citep{2022ApJ...929...31C}. Therefore, $j_\phi^{(2)}$ is in general nonzero when the Alfv\'en wave propagates on curved field lines, and
the production of a fms wave is inevitable.

\begin{figure*}[t]
    \centering
    \includegraphics[width=\textwidth]{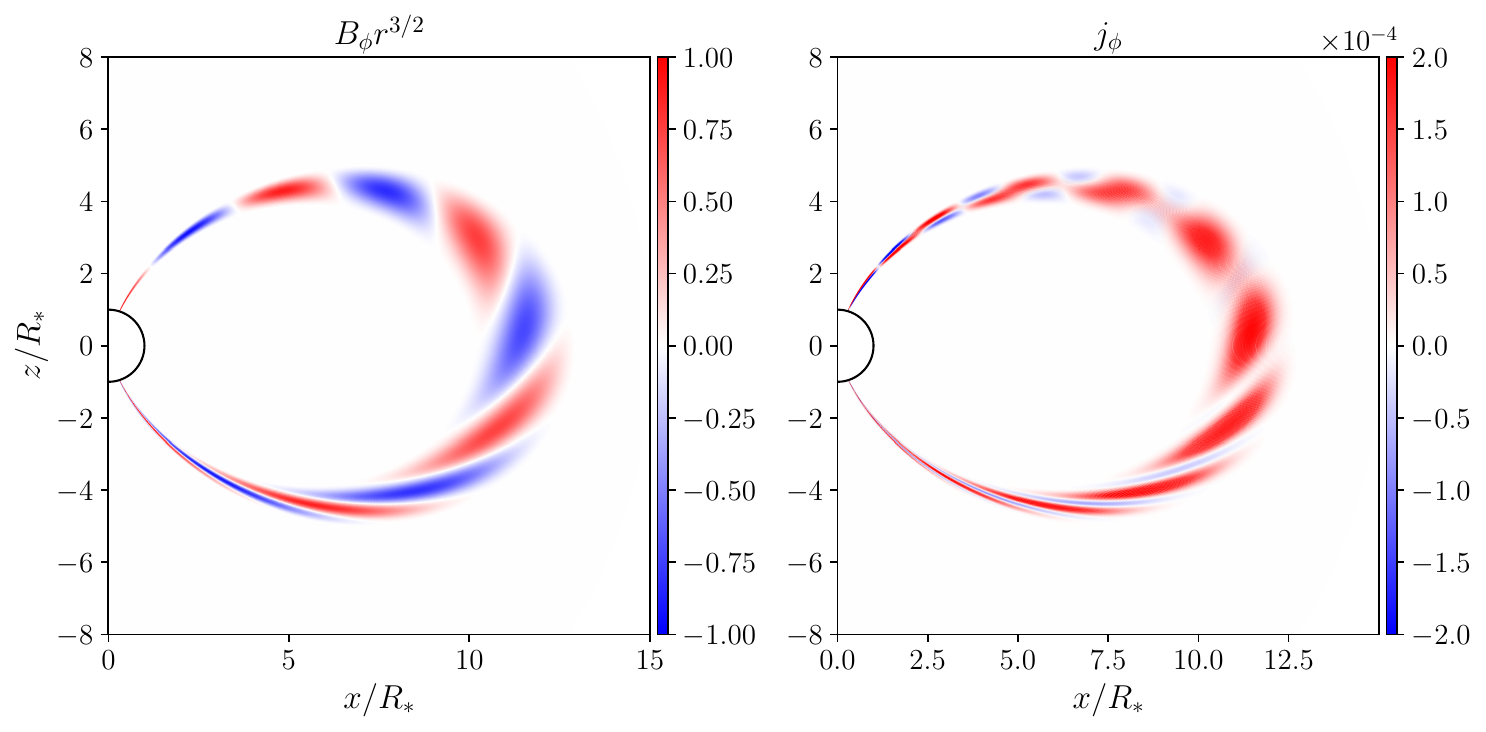}
    \caption{Alfv\'en wave solution computed using the method described in Section~\ref{sec:quant}, evaluated at an arbitrary time $t = 13R_{*}/c$. The wave amplitude is $\delta B/B_{0}\approx 1/40$ at launch point. The wave is launched in the flux tube $10R_{*} \leq p \leq 13R_{*}$, with wavelength $\lambda\approx 6R_{*}$. The solution is interpolated onto a $2000\times 2000$ 2D spherical grid.}
    \label{fig:2d-nonrotating}
\end{figure*}

Next, we want to find the Fourier components of $j_{\phi}^{(2)}$. The real part of Equation~\eqref{eq:Bphi-time-dependent} can be written as:
\begin{equation}
    B_{\phi}^{(1)} = \hat{f}_{1}\cos\omega t + \hat{f}_{2}\sin\omega t,
\end{equation}
where $\hat{f}_{i} = r^{-3/2}f_{i}$. Plugging this into Equation~\eqref{eq:jphi}, the structure of the first term is:
\begin{equation}
    \label{eq:jphi-fourier}
    \begin{split}
      B_{\phi}^{(1)}\hat{D}_{B}B_{\phi}^{(1)} &= \frac{1}{2}\left[\left(\hat{f}_{1}\hat{D}_{B}\hat{f}_{2} + \hat{f}_{2}\hat{D}_{B}\hat{f}_{1}\right)\sin 2\omega t\right. \\
      & +\left. \left(\hat{f}_{1}\hat{D}_{B}\hat{f}_{1} - \hat{f}_{2}\hat{D}_{B}\hat{f}_{2}\right)\cos 2\omega t\right. \\
      & +\left. \left(\hat{f}_{1}\hat{D}_{B}\hat{f}_{1} + \hat{f}_{2}\hat{D}_{B}\hat{f}_{2}\right)\right].
    \end{split}
\end{equation}
The second term in Equation~\eqref{eq:jphi} has a similar structure. Therefore, $j_{\phi}^{(2)}$ consists of an oscillating component with frequency $2\omega$, as well as a non-oscillating component. The oscillating component sources an outgoing fms wave with frequency $2\omega$, as discussed in Section~\ref{sec:formalism}. The non-oscillating component does not directly source an outgoing wave, but as the primary Alfv\'en wave bounces back and forth on a closed field line tube, this component will lead to oscillations at a frequency comparable to $c/L$, where $L$ is the length of the field line. This low frequency oscillation may source an additional fms wave, as is seen in simulations performed by~Bernardi et al (in prep).

To compute the concrete radiation power in fms waves, we specialize to a particular magnetospheric configuration. In a static, axisymmetric dipole magnetosphere, an Alfv\'en wave is launched from the surface in a flux tube between $p_{1}$ and $p_{2}$. The magnitude of the Alfv\'en wave is determined by the boundary conditions at launch point (see Appendix~\ref{sec:app-alfven-solution} for more discussion).
In order for the wave profile to smoothly go to zero at the $p = p_{1,2}$ boundaries, we apply a transverse profile function $a(p) = \cos^{2}(\pi (p - p_{m}) / (p_2 - p_{1}))$, where $p_{m} = (p_{1} + p_{2}) / 2$ marks the approximate middle point of the flux tube.

The Alfv\'en wave is launched from the magnetic footpoint in the northern hemisphere. For simplicity, the boundary at the southern footpoint of the flux tube is assumed to be perfectly absorbing, which allows us to avoid the discussion of wave reflection. We solve Equation~\eqref{eq:f-equation-mu} for 100 evenly distributed field lines inside the flux tube $p_{1} \leq p \leq p_{2}$, then interpolate the results to an evenly spaced grid in $(r, \theta)$. The toroidal current $j_{\phi}^{(2)}$ is computed using a second-order finite difference scheme on the spherical grid. A result of such a calculation is shown in Figure~\ref{fig:2d-nonrotating}.

Once we have computed the spatial dependence of $j_{\phi}^{(2)}$, we can separate the oscillating part and the constant part in Equation~\eqref{eq:jphi-fourier}. The constant part does not produce an electromagnetic wave since we do not treat the reflection of the primary Alfv\'en wave. The oscillatory part directly produces the following vector potential~\citep[see e.g.][]{jackson1999classical}:
\begin{equation}
    \label{eq:Aphi}
    A_{\phi}^{(2)}(\bm{x}) = \frac{1}{c}\int \frac{j_{\phi}^{(2)}(\bm{x}')e^{ikr}}{r}\,d^{3}x',
\end{equation}
where $r = \left|\bm{x} - \bm{x}'\right|$, $k = \omega_{2}/c$, and a $e^{-i\omega_{2} t}$ time dependence is understood for both $j_{\phi}$ and $A_{\phi}$. Then the total power emitted in fms waves can be computed by integrating the outgoing Poynting flux at a large radius $R$:
\begin{equation}
    \label{eq:L-fms}
    \begin{split}
    L_\mathrm{fms} &= \frac{c}{4\pi}\int \left|\bm{E}\times \bm{B}\right|R^2\,d\Omega \\
                    &= \frac{c}{4\pi}\int \left|\omega A_{\phi}^{(2)}\right|^{2}R^{2}\,d\Omega.
    \end{split}
\end{equation}
The square amplitude $\left|A_{\phi}^{(2)}\right|^{2}$ also gives the angular profile of the outgoing fms wave.

\subsection{Scaling}
\label{sec:scaling}

\begin{figure}[h]
    \centering
    \includegraphics[width=0.49\textwidth]{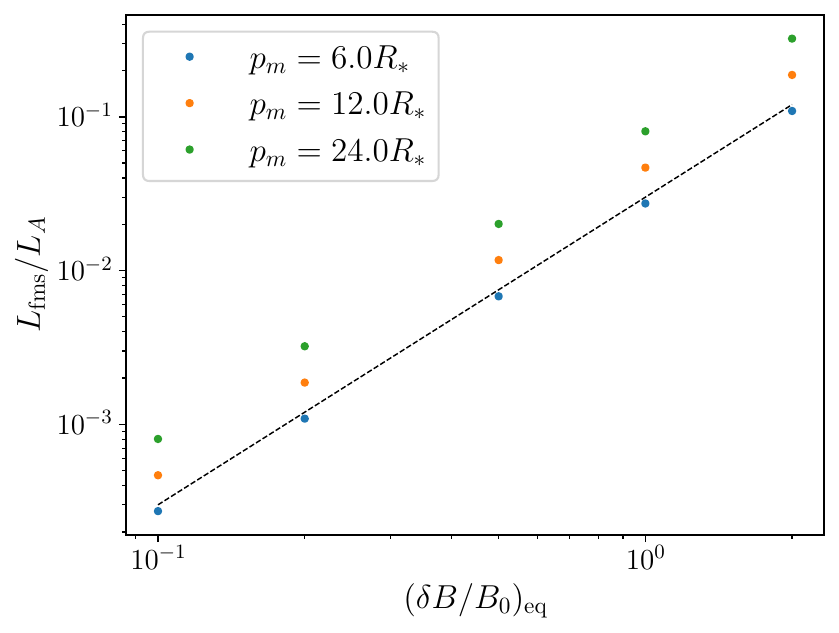}
    \caption{Scaling of the fms wave conversion efficiency with respect to the relative amplitude of the original Alfv\'en wave measured at the equator. Black dashed line corresponds to $L_\mathrm{fms}/L_{A}\propto (\delta B/B_{0})_\mathrm{eq}^{2}$. The conversion efficiency scales quadratically with $(\delta B/B_{0})_\mathrm{eq}$, and scales linearly with the total flux tube length. When varying $p_{m}$, we keep the transverse size of the flux tube $p_{2} - p_{1}$ to be fixed at $2R_{*}$. The Alfv\'en wavelength is taken to be $\lambda_{A} = R_{*}$.}
    \label{fig:scaling-nonrotating}
\end{figure}

We performed this calculation for a series of different $\delta B/B_{0}$ and
$p_{m}$. We keep $p_{2} - p_{1}$ fixed to be $2R_{*}$, so that the transverse
size of the flux tube remains roughly the same. The Alfv\'en wavelength is taken
to be $\lambda_{A} = R_{*}$. The luminosity is computed at a very large radius,
$r = 400R_{*}$, far away from the maximum extent of any of the flux tubes
considered. The results are shown in Figure~\ref{fig:scaling-nonrotating}. The
power in the outgoing fms wave scales exactly quadratically with respect to
$(\delta B/B_{0})_\mathrm{eq}$, which is the relative amplitude of the primary
Alfv\'en wave at at the equator. We measure this quantity at the middle of the
flux tube, $p_{m} = (p_{1} + p_{2})/2$. The fms wave power also scales
approximately linearly with $p_{m}$, which is proportional to the length of the
flux tube.

The quadratic scaling with respect to the relative amplitude of the wave is not
a surprise, as $j_{\phi}^{(2)}$ manifestly depends on $(\delta B^{(1)})^{2}$.
The linear scaling with respect to $p_{m}$ is likely a result of a larger
emitting region size, which scales linearly with $p_{m}$. Both of these features
agree with what was reported in FFE simulations~\citep{2021ApJ...908..176Y}. The
full numerical simulations include effects beyond second order, as well as the
reflection of the primary Alfv\'en wave and its self-intersection. However, it
was shown that most of the $\omega_{2} = 2\omega$ fms wave production happens
during the first passing of the Alfv\'en wave, therefore we expect that the
calculation here is a relatively good approximation of the full nonlinear
behavior until $(\delta B/B_{0})_\mathrm{eq} \gg 1$.

The reason that outgoing fms wave power significantly reduces after reflection
is due to the dephasing of the wave. As the Alfv\'en wave propagates on the
curved flux tube, the path length of the inner edge of the flux tube is shorter
than the outer edge. The wave therefore builds up a phase difference on adjacent
field lines. This effect can be clearly seen in Figure~\ref{fig:2d-nonrotating}.
Since $A_{\phi}$ is the integral of $j_{\phi}$ with an oscillatory function
$e^{ikr}$, the contribution becomes small when the variation scale of $j_{\phi}$
becomes much smaller than $k = \omega/c$. The amount of dephasing is especially
significant for extended field lines~\citep{2022ApJ...929...31C}, and we expect
the only effective window for wave conversion via this channel is during the
initial passing of the Alfv\'en wave in the flux tube, before significant
dephasing has happened.

The same formalism outlined in this section can be applied to multi-wave
interactions. It is well-known that in the force-free limit, $A+A\rightarrow F$
is an allowed nonlinear conversion channel~\citep{1998PhRvD..57.3219T}. In the
formalism developed in this paper, one can include this effect by simply
considering $B_{\phi}^{(1)}$ as a sum of multiple wave components,
$B_{\phi}^{(1)} = B_{\phi,a}^{(1)} + B_{\phi,b}^{(1)}$, and calculate the
toroidal current using this sum. The resulting current will have oscillating
components with frequencies $\omega = \omega_{a} \pm \omega_{b}$, $2\omega_{a}$,
and $2\omega_{b}$, in addition to the bouncing frequency associated with the
length of the flux tube. The outgoing fms wave luminosity will scale with the
product of the relative amplitudes.

The energy of the outgoing fms wave comes at the expense of the original
Alfv\'en wave. The second order Alfv\'en component can account for this amplitude
decrease. However, full energy conservation including the second order fms wave
is difficult to incorporate into the calculation. The energy of the second order
fms wave shows up at fourth order in the energy expansion, which is at the same
order as terms such as $\bm{B}^{(0)}\cdot \bm{B}^{(4)}$ or
$\bm{B}^{(1)}\cdot \bm{B}^{(3)}$. Since we terminate the asymptotic expansion at
the second order, treating the issue of energy conservation is beyond the scope
of this paper.

\section{Closed Zone of a Rotating Magnetosphere} \label{sec:rotating}

In a rotating magnetosphere, the background electric field $\bm{E}^{(0)}$ is
no longer zero. Instead,
$\bm{E}^{(0)} = -(\bm{v}/c)\times \bm{B}^{(0)}$. As a result, the
asymptotic expansion discussed in Section~\ref{sec:non-rotating} needs to be
modified. In this section, we focus on the limit of slow rotation, and employ a
two-scale asymptotic expansion in both $\delta B/B_{0}$ and the rotation speed
$v_{\phi}/c$. The first expansion is identical to what we used in the previous
section, and we continue to use superscript $(1)$ to denote the first order
terms. We will use an additional superscript $[1]$ to denote the first order
corrections due to the rotation of the star. The rotation-induced background
electric field is written as $\bm{E}^{(0),[1]}$ in this notation system.

We will focus on the closed zone of the magnetosphere, where no poloidal current
is flowing. The open zone in the force-free magnetosphere will have nonzero
magnetospheric current flowing to infinity, and a nonzero $B_{\phi}^{(0),[1]}$
as a result~\citep[see e.g.][]{1999ApJ...511..351C}. For slowly rotating neutron
stars such as most magnetars discovered so far, the closed zone dominates the region near
the star, therefore the results of this section should apply to most of the
Alfv\'en waves launched from the surface.

Under the two-scale expansion, the $(0)$-th order current density acquires a
second-order correction in $v_{\phi}/c$:
\begin{equation}
    \label{eq:j0-rotating}
    \begin{split}
      \bm{j}^{(0),[1]} &= 0, \\
      \bm{j}^{(0),[2]} &= -\frac{c}{4\pi}\nabla\cdot \bm{E}^{(0),[1]} \frac{\bm{E}^{(0),[1]}\times \bm{B}^{(0),[0]}}{B_{0}^{2}} \\
                             & = \rho_\mathrm{GJ}v_{\phi}\hat{\boldsymbol{\phi}},
    \end{split}
\end{equation}
where $\rho_\mathrm{GJ}$ is the Goldreich-Julian charge
density~\citep{1969ApJ...157..869G}. The current is second order because both $v_{\phi}$ and $\rho_\mathrm{GJ}$ are first order in rotation. This second order current sources a change
in $B^{(0)}$ leading to a deviation from the non-rotating dipole solution in the
steady state in the closed zone:
\begin{equation}
    \label{eq:B0-rotating}
    \nabla\times \bm{B}^{(0),[2]} = 4\pi \bm{j}^{(0),[2]}.
\end{equation}
There is no correction to the background magnetic field at $[1]$-st order in the
closed zone. Therefore, to first order in $v_{\phi}/c$, the magnetic field lines
remain dipole-shaped, and we can continue using $(q, p, \phi)$ as our field-aligned
coordinate system.

The non-rotating Alfv\'en wave solution we obtained in
Section~\ref{sec:non-rotating} is now denoted by $\bm{B}^{(1),[0]}$ and
$\bm{E}^{(1),[0]}$. The equations for the first order Alfv\'en wave remain
unchanged:
\begin{equation}
  \label{eq:eq-alfven-1-0}
  \begin{split}
  \partial_{t} \bm{B}_{A}^{(1),[0]} &= -c\nabla\times \bm{E}_{A}^{(1),[0]} \\
  \partial_{t} \bm{E}_{A}^{(1),[0]} &= c\nabla\times \bm{B}_{A}^{(1),[0]} - 4\pi \bm{j}^{(1),[0]},
  \end{split}
\end{equation}
where $\bm{j}^{(1),[0]}$ is given by Equation~\eqref{eq:j1}. We are now
interested in the nonlinear effects at first order in rotation, assuming that a
single Alfv\'en wave described by Equation~\eqref{eq:B-ansatz-f} is injected
into the magnetosphere.

Since
$\bm{E}^{(0),[1]} = -(\bm{v}/c)\times \bm{B}_{0} \propto \hat{\bm{p}}$,
under axisymmetry,
$\nabla\times \bm{E}^{(0),[1]} \sim \partial_{q}E_{p}^{(0),[1]}\hat{\boldsymbol{\phi}}$.
Similarly we have already established in the previous section that
$\nabla\times \bm{E}^{(1),[0]} \sim \partial_{q}E_{p}^{(1),[0]}\hat{\boldsymbol{\phi}}$.
Therefore, both $\bm{E}^{(0),[1]}\cdot \nabla\times \bm{E}^{(1),[0]}$
and $\bm{E}^{(1),[0]}\cdot\nabla \times \bm{E}^{(0),[1]}$ are zero. As a result, the nonzero terms in the 1st order correction to $\bm{j}^{(1)}$ are:
\begin{equation}
    \label{eq:j1-rotating}
    \begin{split}
      \bm{j}^{(1),[1]} &= \frac{c}{4\pi}\left[\frac{\bm{B}^{(0),[0]}\cdot \nabla\times \bm{B}^{(1),[1]}}{B_{0}^{2}}\bm{B}^{(0),[0]} \right. \\
                          &\quad - \nabla\cdot \bm{E}^{(1),[0]}\frac{\bm{E}^{(0),[1]}\times \bm{B}^{(0),[0]}}{B_{0}^{2}} \\
      &\quad - \left.\nabla\cdot \bm{E}^{(0),[1]}\frac{\bm{E}^{(1),[0]}\times \bm{B}^{(0),[0]}}{B_{0}^{2}}\right].
    \end{split}
\end{equation}
The structure of $\bm{j}^{(1),[1]}$ is very similar to the second order current
in the non-rotating case, given in Equation~\eqref{eq:j2}. The first term is
formally identical to the $[0]$-th order current~\eqref{eq:j1}, and represents
the current associated with the $[1]$-st order corrected Alfv\'en wave that
flows along the background field line. The remaining two terms are both along the
$\hat{\boldsymbol{\phi}}$ direction, and we group them as
$\bm{j}_{\phi}^{(1),[1]}$.

Since $\bm{j}_{\phi}^{(1),[1]}$ does not depend on $\bm{E}^{(1),[1]}$ or
$\bm{B}^{{(1),[1]}}$, the Maxwell equations for the $[1]$-st order electric and
magnetic fields sourced by this current are identical to the vacuum, linear version:
\begin{equation}
    \label{eq:E-B-1-1}
    \begin{split}
    \partial_{t} \bm{B}_{F}^{(1),[1]} &= -c\nabla \times \bm{E}_{F}^{(1),[1]} \\
    \partial_{t} \bm{E}_{F}^{(1),[1]} &= c\nabla \times \bm{B}_{F}^{(1),[1]} - 4\pi \bm{j}_{\phi}^{(1),[1]}.
    \end{split}
\end{equation}
Therefore, the $[1]$-st order toroidal current again acts as an antenna that
radiates fast waves which is identical to vacuum electromagnetic waves in the
FFE limit (albeit with a particular polarization). We added subscript $F$ to
$\bm{E}$ and $\bm{B}$ to distinguish it from the corrections to the Alfv\'en
wave at this order. Since $\bm{j}_{\phi}^{(1),[1]}$ has only one factor of
$\bm{E}^{(1),[0]}$, its frequency will be the same as the Alfv\'en wave
frequency, in contrast with the nonrotating case where the fast wave at $(2)$-nd
order has double the Alfv\'en wave frequency. Its amplitude scales as
$O(\delta B/B_{0})O(v_{\phi}/c)$, which is different from the non-rotating case.
At very small $\delta B/B_{0}$, this first order fast wave dominates the
Alfv\'en to fast wave conversion, while at larger amplitudes the second order
non-rotating contribution will dominate. This is consistent with the force-free
simulations conducted by~\citet{2021ApJ...908..176Y}.

\begin{figure}[h]
    \centering
    \includegraphics[width=0.5\textwidth]{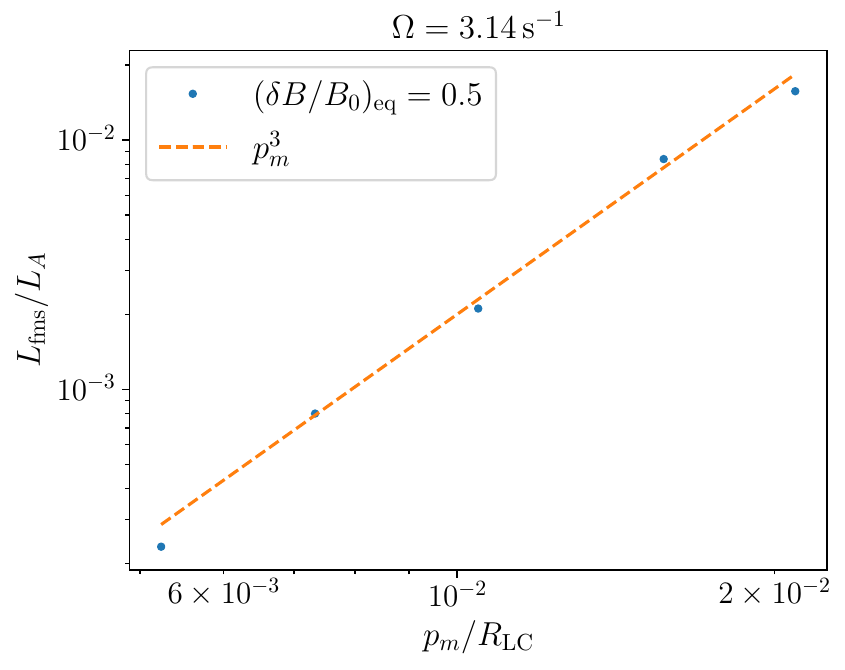}
    \caption{The Alfv\'en to fms wave conversion efficiency at first order in rotation scales with $p_{m}^{3}$. The calculations are done with fixed $(\delta B/B_0)_\mathrm{eq} = 0.5$ and fixed rotation frequency $\Omega = 3.14\,\mathrm{s}^{-1}$, which corresponds to a rotation period of $2\,\mathrm{s}$. When calculating the current in different flux tubes, we keep $p_{2} - p_{1} = 10R_{*}$ for all cases, so that the transverse size of the flux tube remains roughly the same. The Alfv\'en wave length is taken to be $5R_{*}$. $L_\mathrm{fms}$ is computed at $R=1000R_*$.}
    \label{fig:scaling-rotating}
\end{figure}

The power of fms waves produced in the rotating magnetosphere can be computed in
the same way as described in Section~\ref{sec:quant}.
Figure~\ref{fig:scaling-rotating} shows the scaling of conversion efficiency
with respect to the maximum field line extent $p_{m}$. The result is consistent
with $L_\mathrm{fms}/L_{A}\propto p_{m}^{3}$. This is not surprising, as the fms
wave amplitude scales as $\delta B v_{\phi}$, and its luminosity scales as
$\delta B^{2}v_{\phi}^{2}$. Since the primary Alfv\'en wave luminosity also
scales as $\delta B^{2}$, this factor drops out. On the other hand,
$v_{\phi} = \rho \Omega \sim p_{m}\Omega$, leading to a scaling factor of
$p_{m}^2$. In addition, we expect the fms wave luminosity to scale linearly with
the length of the flux tube, similar to the nonrotating case. Therefore
$L_\mathrm{fms}/L_{A}\propto p_{m}^{3}$ is consistent with analysis presented in
this section. Due to the rather strong scaling with $p_{m}$, we expect
$L_\mathrm{fms}/L_{A}$ to reach order unity near $p_{m}/R_\mathrm{LC}\sim 0.1$.
This fraction may be lower if the duration of the Alfv\'en wave is shorter than
the light-crossing time of the flux tube, but it seems relatively easy to convert a
significant fraction of the Alfv\'en wave energy to fms waves in a rotating
magnetosphere, as long as the primary Alfv\'en wave propagates on field lines that extend to a
fraction of the light cylinder radius.

Note that our main focus is not to find the corresponding first order normal
modes in a rotating magnetosphere. Such a calculation has been carried out
by~\citet{2021ApJ...908..176Y} in their Appendix D, under a WKB approximation
where $kL\gg 1$. In fact, their result includes up to second order terms in
rotation, or order $(1)$, $[2]$ in our language:
$\delta \bm{E} = \bm{E}^{(1),[0]} + \bm{E}^{(1),[1]} + \bm{E}^{(1),[2]}$.
Instead, we keep track of the separate orders and only keep the first order
corrections in rotation, which leads to much simpler equations for each order
without the WKB assumption. In fact, since the $[2]$-nd order corrections to
the background magnetic field is nonzero in the closed zone, it would have been
inappropriate if we extended the dipole-based formalism in this paper to second
order in rotation.

\section{Discussions} \label{sec:discussion}

In this paper, we have solved for the small-amplitude Alfv\'enic normal modes in
a dipole background magnetic field. Using an asymptotic expansion, we
demonstrated that a single Alfv\'en wave will spontaneously convert to fast
magnetosonic waves as it propagates on the curved field line. In a non-rotating
dipole magnetosphere, the resulting fms wave will have frequency twice the
original Alfv\'en wave frequency. In the closed zone of a rotating force-free
magnetosphere, the fms wave first shows up at first order in a two-scale
expansion, and its frequency is the same as the original Alfv\'en wave
frequency. In both cases, the fast wave is sourced by a toroidal electric
current that acts like an antenna. This picture not only elucidates the wave
conversion process, but also provides a concrete means to compute the properties of the
outgoing fms wave using standard electrodynamics techniques, without resorting
to global FFE simulations.

The numerical solution of the Alfv\'en wave profile can be a useful tool in
future research on wave phenomena in the magnetospheres of neutron stars,
especially for low frequency Alfv\'en waves whose wavelengths are comparable to
the global scales. It can also be used as a benchmark tool for testing FFE or
PIC simulation codes in spherical coordinates. Although we do not treat
reflection in this paper, it can be added relatively easily using the usual
trick of launching additional waves with a time delay from outside the domain.
The primary Alfv\'en wave may interact nonlinearly with its reflection, leading
to an enhanced production of fms waves.

This paper mainly concerns the spontaneous evolution of a single Alfv\'en wave
launched from the surface of the neutron star. However, multi-wave interaction
(e.g.\ interaction of the primary Alfv\'en wave with its reflection) can be
included in the same framework too. In particular, multiple Alfv\'en waves
superimpose linearly at the first order, which will lead to a range of frequency
components in the second order current, Equation~\eqref{eq:jphi}. As a result,
fms waves of different frequencies will be sourced, and the calculation is
similar to the single wave case.

The fms wave produced from spontaneous conversion may suffer from its own
nonlinear effects. For example, if the outgoing fms wave is strong enough, it
may steepen within the magnetosphere into a shock at each
wavelength~\citep[see][]{2022arXiv221013506C,2023ApJ...959...34B}. This shock
may produce interesting radiation signals such as strong X-ray emission. If the
fms wave does not steepen within the magnetosphere, it may interact nonlinearly
with the current sheet near the light cylinder, leading to current sheet
compression and enhanced reconnection rates~\citep[see
e.g.][]{2019MNRAS.483.1731L,2022ApJ...932L..20M,2023RAA....23c5010W}. The results of this paper can provide a framework for estimating the properties of the fms waves converted from Alfv\'en waves, potentially contributing to a more complete understanding of X-ray or radio bursts from magnetars.

\begin{acknowledgments}

    We thank Yuanhong Qu and Xinyu Li for helpful discussions and comments on the
    manuscript. AC and YY acknowledge support from NSF grants DMS-2235457 and
    AST-2308111. YY also acknowledges support from the Multimessenger Plasma Physics Center (MPPC), NSF grant PHY-2206608, and support from the Simons Foundation (MP-SCMPS-00001470).

\end{acknowledgments}

\appendix

\section{Dipole Coordinate Identities}
\label{sec:app-dipole-coord}

In this appendix we list some common identities in dipole coordinates that are used in the main text. Much of the content here follows \citet{2006physics...6044S}. First, the dipole coordinates as defined by equations~\eqref{eq:dipole-coord} take the range $q\in (-\infty, \infty)$ and $p\in [0, \infty)$. It is not an orthonormal coordinate system, and the metric tensor is:
\begin{equation}
    \label{eq:dipole-metric}
    g_{ij} = \mathrm{diag}\left(\frac{r^{6}}{\delta^{2}}, \frac{\sin^{6}\theta}{\delta^{2}}, r^{2}\sin^{2}\theta\right),
\end{equation}
where $\delta = \sqrt{1 + 3\cos^{2}\theta}$. Since Alfv\'en waves are launched from the stellar surface, it is convenient to define a field line length $s$ that starts at the stellar radius $r = R_{*}$. This turns out to be nontrivial since the stellar surface does not coincide with a coordinate surface. \citet{2022ApJ...929...31C} wrote down an exact expression for $s$ for a field line starting at $\mu_{0} = \cos\theta_{0}$:
\begin{equation}
    \label{eq:fieldline-length}
    s(p, \mu) = p\left[F(\mu_{0}) - F(\mu)\right],
\end{equation}
where $\mu = \cos\theta$ and
\begin{equation}
    \label{eq:Fmu}
    F(\mu) = \frac{1}{2}\mu \sqrt{1 + 3\mu^{2}} + \frac{\sinh^{-1}(\sqrt{3}\mu)}{2\sqrt{3}}.
\end{equation}
The field line length $s$ can be thought of as a rescaling of the coordinate variable $q$, although constant $s$ surfaces do not coincide with constant $q$ surfaces. Despite the seemingly complex expression, some of its coordinate derivatives are relatively simple:
\begin{equation}
    \label{eq:s-derivatives}
    \frac{\partial s}{\partial q} = \frac{r^{3}}{\delta},\quad \frac{\partial s}{\partial \theta} = p\delta \sin\theta.
\end{equation}

The 3-dimensional curl of a vector field in these coordinates can be written as:
\begin{equation}
    \label{eq:curl-dipole-coord}
    \begin{split}
    \nabla\times \bm{A} &= \hat{\bm{q}}\left[\frac{\delta}{\sin^{3}\theta}\frac{\partial A_{\phi}}{\partial p} + \frac{1 - 3\cos^{2}\theta}{r\delta\sin\theta}A_{\phi} - \frac{1}{r\sin\theta}\frac{\partial A_{p}}{\partial \phi}\right] \\
      &+ \hat{\bm{p}}\left[\frac{1}{r\sin\theta}\frac{\partial A_{q}}{\partial \phi} - \frac{\delta}{r^{3}}\frac{\partial A_{\phi}}{\partial q} + \frac{3\cos\theta}{r\delta}A_{\phi}\right] \\
      &+ \hat{\boldsymbol{\phi}}\left[\frac{\delta}{r^{3}}\frac{\partial A_{p}}{\partial q} - \frac{6\cos\theta}{r\delta^{3}}(1 + \cos^{2}\theta)A_{p} - \frac{\delta}{\sin^{3}\theta}\frac{\partial A_{q}}{\partial p} - \frac{3\sin\theta}{r\delta^{3}}(1 + \cos^{2}\theta)A_{q}\right].
    \end{split}
\end{equation}
The divergence of a 3-dimensional vector field in dipole coordinates is:
\begin{equation}
    \label{eq:div-dipole-coord}
    \begin{split}
      \nabla\cdot \bm{A} &= \frac{\delta^{2}}{r^{6}}\frac{\partial}{\partial q}\left(\frac{r^{3}}{\delta}A_{q}\right) + \frac{\delta^{2}}{\sin^{6}\theta}\frac{\partial}{\partial p}\left(\frac{\sin^{3}\theta}{\delta}A_{p}\right) + \frac{4}{r\delta\sin\theta}A_{p} + \frac{1}{r\sin\theta}\frac{\partial A_{\phi}}{\partial \phi} \\
      &= \frac{\delta}{r^{3}}\frac{\partial A_{q}}{\partial q} - \frac{3\cos\theta}{r\delta^{3}}(3 + 5\cos^{2}\theta)A_{q} + \frac{\delta}{\sin^{3}\theta}\frac{\partial A_{p}}{\partial p} + \frac{4}{r\delta^{3}\sin\theta}(1 - 3\cos^{4}\theta)A_{p} + \frac{1}{r\sin\theta}\frac{\partial A_{\phi}}{\partial \phi}
    \end{split}
\end{equation}

\section{Alfv\'en Wave Solutions}
\label{sec:app-alfven-solution}

In this appendix we outline our method to obtain solutions to Equation~\eqref{eq:f-equation-mu} that represents a propagating wave. This is a second order ODE with no regular singularity points, and its solutions are not described by well-known functions such as hypergeometric functions. We resort to numerical techniques to solve this equation.

In general, a second order linear ODE admits two independent solutions, and the general solution is a linear combination of them:
\begin{equation}
    f(\mu) = C_{1}f_{1}(\mu) + C_{2}f_{2}(\mu).
\end{equation}
Since we are interested in wave-like solutions, we take the linear combinations
$f_{1}\pm if_{2}$, where the sign corresponds to right-going or left-going waves. We demand the following boundary conditions which mimic $\cos kx$ and $\sin kx$:
\begin{equation}
    \label{eq:f12-boundary-condition}
    \begin{split}
      f_{1}(\mu_{0}) &= A,\quad f_{1}'(\mu_{0}) = 0, \\
      f_{2}(\mu_{0}) &= 0,\quad f_{2}'(\mu_{0}) = K,
    \end{split}
\end{equation}
where $A$ is the overall amplitude of the wave, and $K$ is a constant that needs to be chosen so that $f_{1}$ and $f_{2}$ are consistent with a traveling wave. We fix the constant by demanding $E_{p} = -B_{\phi}$ at the stellar surface, which is equivalent to approximating the solution as a plane wave at the launch point. Although this is an approximation, direct FFE simulations have shown that this assumption is reasonably accurate at the launch point~(Bernardi et al. in prep.), but the equality typically ceases hold after the wave has propagated for some distance.

\begin{figure}[h]
    \centering
    \includegraphics[width=0.5\textwidth]{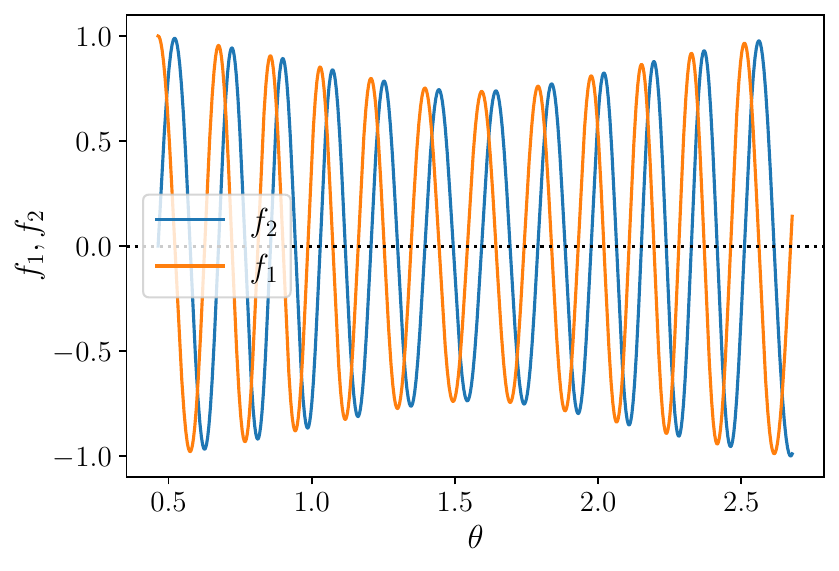}
    \caption{Independent Alfv\'en wave solutions $f_{1}$ and $f_{2}$, normalized such that the maximum amplitude is $A = 1$. The solutions are obtained on the field line $p = 5R_{*}$, and the wave is launched at frequency $\omega = 2\pi c/R_{*}$. $f_{1}$ is the analog of $\cos kx$ and $f_{2}$ is the analog of $\sin kx$.}
    \label{fig:f1-f2}
\end{figure}

In order to determine $K$ using the condition $E_{p} = -B_{\phi}$ at the launch point, we write down the expressions for $E_{p}$ and $B_{\phi}$ using Equation~\eqref{eq:Ep} and $\partial s/\partial q = r^{3}/\delta$:
\begin{equation}
    B_{\phi} = r^{-3/2}(f_{1} + if_{2})e^{-i\omega t},\quad E_{p} = -\frac{ic}{\omega}r^{-3/2}(\partial_{s}f_{1} + i\partial_{s}f_{2}) e^{-i\omega t}.
\end{equation}
Therefore, the boundary condition we apply is simply:
\begin{equation}
    f_{1}(s = 0) = -\frac{c}{\omega}\partial_{s}f_{2}(s = 0)\Longrightarrow K = \frac{\omega p\delta}{c} A.
\end{equation}
The resulting solutions $f_{1}$ and $f_{2}$ are shown in Figure~\ref{fig:f1-f2}. Note that these two solutions are completely determined by the boundary conditions~\eqref{eq:f12-boundary-condition} at the left end, therefore the right boundary conditions are left unconstrained. This choice is to mimic the launching of Alfv\'en waves from crustal motion at one end. When the wave arrives at the opposite end, it will reflect off the stellar surface, and the full solution will be a linear superposition of the original wave plus an additional incoming wave launched from the opposite footpoint.

\end{document}